\documentstyle[times]{article}
\begin{document}

\title{Stationary Electromagnetic Fields of a Slowly
Rotating Magnetized Neutron Star in General Relativity}

\author{L.~Rezzolla$^{1,2}$, B.~J.~Ahmedov$^{3,4}$, and
	J.~C.~Miller$^{1,5}$
	\\ \\ \\
	{\small {\it $^{1}$SISSA, International School 
	for Advanced Studies,}}
	{\small {\it Via Beirut 2-4, 34013 Trieste, Italy}}
	\\ 
	{\small {\it $^{2}$INFN, Department of Physics, University of
        Trieste, Via Valerio, 2 34127 Trieste, Italy}}
	\\
	{\small {\it $^{3}$Institute of Nuclear Physics,
	Ulughbek, Tashkent 702132, Uzbekistan}}
	\\
	{\small {\it $^{4}$AS-ICTP, The Abdus Salam
	International Center for Theoretical Physics, Strada
	Costiera, 11 34014 Trieste, Italy}}
	\\ 
	{\small {\it $^{5}$Nuclear and Astrophysics Laboratory,
	University of Oxford,}}
	{\small {\it Keble Road, Oxford OX1 3RH}}}

\maketitle

\bigskip
\bigskip
\bigskip
\begin{abstract}
Following the general formalism presented by Rezzolla,
Ahmedov and Miller$\footnotesize{^{\cite{ram00}}}$, we
here derive analytic solutions of the electromagnetic
fields equations in the internal and external background
spacetime of a slowly rotating highly conducting
magnetized neutron star. The star is assumed to be
isolated and in vacuum, with a dipolar magnetic field not
aligned with the axis of rotation. Our results indicate
that the electromagnetic fields of a slowly rotating
neutron star are modified by general relativistic effects
arising from both the monopolar and the dipolar parts of
the gravitational field. The results presented here
differ from the ones discussed by Rezzolla, Ahmedov and
Miller$\footnotesize{^{\cite{ram00}}}$ mainly in that we
here consider the interior magnetic field to be dipolar
with the same radial dependence as the external
one. While this assumption might not be a realistic one,
it should be seen as the application of our formalism to
a case often discussed in the literature.
\end{abstract}

\vskip 2.0 truecm
\section{INTRODUCTION}
\vskip 0.5 truecm

	The rich phenomenology observed from pulsars has
long  motivated the study of the electromagnetic
fields interior and exterior to a rotating neutron
star. Furthermore, since neutron stars are among the most
relativistic stellar objects, the study of
electromagnetic fields in strongly curved spacetimes has
been the subject of past and recent interest. Indeed, the
general relativistic modifications to the solutions of
the Maxwell equations in the curved spacetime of a
relativistic star have been studied since the 1960s. The
initial work of Ginzburg and
Ozernoy$\footnotesize{^{\cite{go64}}}$, Anderson and
Cohen$\footnotesize{^{\cite{ac70}}}$ and of
Petterson$\footnotesize{^{\cite{p74}}}$ on the stationary
electromagnetic fields in a Schwarzschild external
background geometry have revealed that, in general, the
spacetime curvature amplifies the magnetic fields outside
compact magnetic stars. More recent work by Geppert, Page
and Zannias$\footnotesize{^{\cite{gpz00}}}$ has
reconsidered this problem also in the context of
non-stationary magnetic fields.

	The general relativistic effects induced by
rotation of the star were first investigated by Muslimov
and Tsygan$\footnotesize{^{\cite{mt92}}}$ who considered
the case of a slowly rotating star surrounded by vacuum
and with a prescribed dipole magnetic field aligned with
the rotation axis. Muslimov and
Harding$\footnotesize{^{\cite{mh97}}}$ then extended this
to treat a non-vacuum exterior. More recently, Rezzolla,
Ahmedov and Miller$\footnotesize{^{\cite{ram00}}}$ have
considered stationary and non-stationary solutions of
the Maxwell equations in the internal and external
background spacetime of a slowly rotating magnetized
relativistic star surrounded by vacuum and with a dipole
magnetic field not aligned with the axis of rotation. In
an independent study, Konno and
Kojima$\footnotesize{^{\cite{kk00}}}$ have also
considered a similar scenario but have limited their
investigation to the stationary solution of an aligned
dipole.

	In this paper we make use of the general
expressions formulated by Rezzolla et
al.$\footnotesize{^{\cite{ram00}}}$ to study the case,
recently discussed in the
literature$\footnotesize{^{\cite{kk00}}}$, in which the
magnetic field interior to the star is dipolar and has
the same radial dependence as the magnetic field outside
it. While this is strictly speaking incorrect since the
interior magnetic field cannot be determined
independently of the structure of the rotating star, the
error made with this assumption is small and is
compensated by the possibility of dealing with a simple
but instructive case.

	The paper is organized as follows: in Section
\ref{meq} we write the general relativistic equations for
the electromagnetic fields in the background metric of a
slowly rotating star in the form they assume in the
orthonormal tetrads of a family of zero angular momentum
observers.  In Section \ref{srst} we find the stationary
general relativistic solutions to the Maxwell equations
in the spacetime internal and external to the rotating
star. In Section \ref{conclusion} we summarize our
conclusions.

	Throughout, we use a space-like signature
$(-,+,+,+)$ and geometrical units where $c= G = 1$
(However, for those expressions with an astrophysical
application we make exceptions by writing the speed of
light explicitly.). Greek indices run from 0 to 3 and
Latin indices from 1 to 3; semi-colons denote covariant
derivatives, while commas denote partial derivatives.

\section{THE MAXWELL EQUATIONS IN A 
SLOWLY ROTATING SPACETIME}
\label{meq}
\vskip 0.5 truecm

	Hereafter we will assume that the contribution of
the electromagnetic energy density to the total mass
density is negligibly small, even for very highly
magnetized stars. Under this approximation we can solve
the general relativistic Maxwell equations on a given
fixed curved background, rather than the coupled
Einstein-Maxwell equations. In particular, we will
consider the background metric as being that of a
stationary, axially symmetric star approximated to first
order in the angular velocity $\Omega$. Using a
Boyer-Lindquist coordinate system $(t,r,\theta,\phi)$
this metric takes the
form$\footnotesize{^{\cite{h67,ht68,ll71}}}$
\begin{equation}
\label{slow_rot}
ds^2 = -e^{2 \Phi(r)} dt^2 + e^{2
	\Lambda(r)}dr^2 - 2 \omega (r) r^2\sin^2\theta dt d\phi +
	r^2 d\theta ^2+ r^2\sin^2\theta d\phi ^2 \ ,
\end{equation}
where the differential rotation $\omega(r)$ is caused by
frame-dragging and can be interpreted as the
angular velocity of a free falling (inertial) frame. The
radial dependence of $\omega$ in the region of spacetime
internal to the star has to be found as the solution of
the differential equation
\begin{equation}
\label{omg_r_int}
\frac{1}{r^3} \frac{d}{dr}
	\left(r^4 {\bar j} \frac{d {\bar \omega}}{dr}\right)
	+ 4 \frac{d {\bar j}}{dr} {\bar \omega} = 0 \ ,
\end{equation}
where we have defined
\begin{equation}
\label{jbar}
{\bar j} \equiv e^{-(\Phi + \Lambda)} \ ,
\end{equation}
and where
\begin{equation}
\label{omegabar}
{\bar\omega} \equiv \Omega -\omega \ ,
\end{equation}
is the angular velocity of the fluid as measured from the
local free falling (inertial) frame. In the vacuum region
of spacetime external to the star, on the other hand,
$\omega(r)$ is given by the simple algebraic expression
\begin{equation}
\label{omg_r_ext}
\omega (r)\equiv \frac{d\phi}{dt}=-\frac{g_{0\phi}}{g_{\phi\phi}}=
	\frac{2J}{r^3} \ ,
\end{equation}
where $J=I(M,R)\Omega$ is the total angular moment of
metric source as measured from infinity and $I(M,R)$ its
momentum of inertia (see $\footnotesize{^{\cite{m77}}}$
for a discussion of $I$ and its numerical
calculation). Outside the star, the metric
(\ref{slow_rot}) is completely known and explicit
expressions for the other metric functions are given by
\begin{equation}
e^{2 \Phi(r)} \equiv \left(1-\frac{2 M}{r}\right)
	= e^{-2 \Lambda(r)} \ , \hskip 1.0cm
	r > R \ ,
\end{equation}
where $M$ and $R$ are the mass and radius of the star as
measured from infinity.

	Note that the use of a vacuum Schwarzschild
metric in place of
(\ref{slow_rot})$\footnotesize{^{\cite{s95,s97}}}$ should
not be considered satisfactory since the Schwarzschild
metric is intrinsically inadequate for describing
physical systems such as pulsars in which the coupling of
the magnetic field and rotation plays a crucial role in
the production of the observable electric field.

	We start our analysis by recalling the general
relativistic form of the Maxwell equations
\begin{eqnarray}
\label{maxwell_firstpair}
&& 3! F_{[\alpha \beta, \gamma]} =  2 \left(F_{\alpha \beta, \gamma }
	+ F_{\gamma \alpha, \beta} + F_{\beta \gamma,\alpha}
	\right) = 0 \ , \nonumber\\ \\
\label{maxwell_secondpair}
&& F^{\alpha \beta}_{\ \ \ \ ;\beta} = 4\pi J^{\alpha}\ ,
\end{eqnarray}
where the four-current $J^{\alpha}$ can be decomposed
into a conduction current density $j^\alpha$ and a
proper charge current density $\rho_e w^{\alpha}$
\begin{equation}
J^{\alpha}=\rho_e w^\alpha + j^\alpha \ ,
	\hskip 2.0cm j^\alpha w_\alpha \equiv 0 \ ,
\end{equation}
with $w^{\alpha}$ being the components of the conductor's
four-velocity.  If the conduction current is carried by
electrons\footnote{This is a reasonable assumption if the
neutron star has a temperature such that the atomic
nuclei are frozen into a lattice and the electrons form a
completely relativistic, and degenerate gas.}  with
electrical conductivity $\sigma$, Ohm's law can then be
written in its special relativistic form
\begin{equation}
\label{ohm}
j_\alpha = \sigma F_{ \alpha \beta}w^\beta \ ,
\end{equation}
(A general relativistic expression can be found in
$\footnotesize{^{\cite{a99}}}$.).

	The electromagnetic field tensor $F_{\alpha
\beta}$ can be expressed through the electric and
magnetic four-vector fields $E^{\alpha},\ B^{\alpha}$
measured by an observer with four-velocity
$u^{\alpha}$~$\footnotesize{^{\cite{l67,e73}}}$
\begin{equation}
\label{fab_def}
F_{\alpha\beta} \equiv 2 u_{[\alpha} E_{\beta]} +
	\eta_{\alpha\beta\gamma\delta}u^\gamma B^\delta \ ,
\end{equation}
where $T_{[\alpha \beta]} \equiv \frac{1}{2}(T_{\alpha
\beta} - T_{\beta \alpha})$ and
$\eta_{\alpha\beta\gamma\delta}$ is the pseudo-tensorial
expression for the Levi-Civita symbol $\epsilon_{\alpha
\beta \gamma \delta}$~$\footnotesize{^{\cite{s90}}}$
\begin{equation}
\eta^{\alpha\beta\gamma\delta}=-\frac{1}{\sqrt{-g}}
	\epsilon_{\alpha\beta\gamma\delta} \ ,
	\hskip 2.0cm
\eta_{\alpha\beta\gamma\delta}=
	\sqrt{-g}\epsilon_{\alpha\beta\gamma\delta} \ ,
\end{equation}
with $g\equiv {\rm
det}|g_{\alpha\beta}|=-e^{2(\Phi+\Lambda)} r^4
\sin^2\theta$ for the metric (\ref{slow_rot}).

	It will be useful to consider the class of ``zero
angular momentum observers'' or
ZAMOs$\footnotesize{^{\cite{bpt72}}}$.  These are
observers that are locally stationary (i.e. at fixed
values of $r$ and $\theta$) but who are ``dragged'' into
differential rotation with respect to a reference frame
fixed with respect to distant observers. At first order
in $\Omega$ they have four-velocity components given by
\begin{equation}
\label{uzamos}
(u^{\alpha})_{_{\rm ZAMO}}\equiv
	e^{-\Phi(r)}\bigg(1,0,0,\omega\bigg) \ ;
	\hskip 2.0cm
(u_{\alpha})_{_{\rm ZAMO}}\equiv
	e^{\Phi(r)}\bigg(- 1,0,0,0 \bigg) \ .
\end{equation}

	We can now rewrite the Maxwell equations in the
ZAMO reference frame by projecting the electromagnetic
vector fields onto a locally orthonormal tetrad and
indicating their components with ``hatted'' indices: i.e.
\begin{equation}
\label{max1a}
\sin\theta \left(r^2B^{\hat r}\right)_{,r}+
	e^{\Lambda}r\left(\sin\theta B^{\hat \theta}\right)_{,\theta} +
	e^{\Lambda} r B^{\hat \phi}_{\ , \phi} = 0 \ ,
\end{equation}
\begin{eqnarray}
\label{max1b}
\left({r\sin\theta}\right)\frac{\partial B^{\hat r}}{\partial t}
	& = & {e^\Phi} \left[E^{\hat\theta}_{\ ,\phi}- \left(\sin\theta
	E^{\hat \phi} \right)_{,\theta}\right]
	- \left({{\omega} r\sin\theta}\right)B^{\hat r}_{\ ,\phi} \ ,
\\
\label{max1c}
\left({e^{\Lambda}r\sin\theta}\right)
	\frac{\partial B^{\hat \theta}}{\partial t}
	&=& -e^{\Phi+\Lambda} E^{\hat r}_{\ ,\phi} +
	\sin\theta \left(r e^{\Phi} E^{\hat \phi} \right)_{,r}
	- \left({{\omega}e^{\Lambda}r\sin\theta}\right)
	B^{\hat \theta}_{\ ,\phi} \ ,
\\
\label{max1d}
\left({e^{\Lambda}r}\right)
	\frac{\partial B^{\hat \phi}}{\partial t}
	&=& - \left(r e^{\Phi} E^{\hat \theta}\right)_{,r}
	+ e^{\Phi+\Lambda}E^{\hat r}_{ \ ,\theta}
%	\nonumber\\
%	&&
	+ {\sin\theta}\left({\omega} r^2 B^{\hat r}\right)_{,r}
	+ {\omega} e^{\Lambda}r\left(\sin\theta
	B^{\hat \theta}\right)_{,\theta} \
\end{eqnarray}

\noindent and
\begin{eqnarray}
\label{max2a}
&&\sin\theta\left(r^2 E^{\hat r} \right)_{,r}+
	{e^{\Lambda}r}\left(\sin\theta E^{\hat \theta}\right)_{,\theta}
	+ e^{\Lambda}r E^{\hat \phi}_{\;,\phi}
	 =  {4\pi e^{\Lambda}}r^2\sin\theta J^{\hat t}\ ,
\\
\label{max2b}
&&e^{\Phi}\left[\left(\sin\theta  B^{\hat \phi} \right)_{,\theta}
	- B^{\hat\theta}_{\ ,\phi}\right]-
	\left({{\omega} r\sin\theta }\right)E^{\hat r}_{\;,\phi}
	  = \left({r\sin\theta}\right)
	\frac{\partial E^{\hat r}}{\partial t}
	+{4\pi}e^{\Phi}r\sin\theta J^{\hat r} \ ,
\\
\label{max2c}
&&e^{\Phi+\Lambda}B^{\hat r}_{\ ,\phi} - \sin\theta \left(r \ e^{\Phi}
	B^{\hat \phi} \right)_{,r}
	-\left({{\omega} e^{\Lambda}r\sin\theta}\right)
	E^{\hat \theta}_{\;,\phi}
%	 \nonumber\\
%	 && \hskip 3cm
	  =  \left({e^{\Lambda}r\sin\theta}\right)
	\frac{\partial E^{\hat\theta}}{\partial t}
	+{4\pi e^{\Phi+\Lambda}}r\sin\theta J^{\hat\theta} \ ,
\\
\label{max2d}
&&\left(e^{\Phi}r B^{\hat \theta} \right)_{,r} - e^{\Phi+\Lambda}
	B^{\hat r}_{\ ,\theta} + {\sin\theta}\left({\omega}
	r^2E^{\hat r}\right)_{,r} +
	{{\omega} e^{\Lambda}r}
	\left(\sin\theta E^{\hat \theta}\right)_{,\theta}
	 \nonumber\\
	 && \hskip 2.5cm
	 = \left({e^{\Lambda}r}\right)
	\frac{\partial E^{\hat\phi}}{\partial t}
	+{4\pi e^{\Phi+\Lambda}}rJ^{\hat\phi}
	+{4\pi e^{\Lambda}}\omega r^2\sin\theta J^{\hat t} \ .
\end{eqnarray}
Note that, apart from the general relativistic
corrections produced by the static part of gravitational
field and proportional to the metric functions $\Lambda$ and
$\Phi$, the Maxwell equations (\ref{max1a})--(\ref{max2d})
contain new general relativistic terms due to the
coupling between the frame-dragging effect and the
electromagnetic fields. This coupling can then act as a
source of electric and magnetic fields.

	Taking our conductor to be the star with
four-velocity components
\begin{equation}
\label{vel}
w^\alpha \equiv e^{-\Phi(r)}
	\bigg(1,0,0,{\Omega}\bigg) \ ,
\hskip 2.0cm
w_\alpha \equiv
	e^{\Phi(r)} \bigg(-1,0,0,
	\frac{\bar{\omega} r^2\sin ^2\theta}{e^{2\Phi(r)} }\bigg)\ ,
\end{equation}
we can use Ohm's law (\ref{ohm}) to derive the following
explicit components of $J^{\hat \alpha}$ in the ZAMO
frame
\begin{eqnarray}
\label{current1}
&&J^{\hat t} = {\rho_e}+
	\sigma\frac{\bar{\omega}r\sin\theta}
	{e^{\Phi}}E^{\hat\phi}\ ,
\\\nonumber\\
\label{current2}
&&J^{\hat r} = \sigma
	\left( E^{\hat r} -
	\frac{\bar{\omega} r \sin\theta }{e^{\Phi}}
	B^{\hat \theta}\right)\ ,
\\\nonumber\\
\label{current3}
&&J^{\hat\theta} = \sigma\left(E^{\hat \theta}+
	\frac{\bar{\omega} r \sin\theta }{e^{\Phi}}
	B^{\hat r}\right)\ ,
\\\nonumber\\
\label{current4}
&&J^{\hat\phi} = {\sigma E^{\hat \phi}} +
	\frac{\bar{\omega}r\sin\theta}{e^{\Phi}}\rho_e\ .
\end{eqnarray}

	To simplify the problem, hereafter we will adopt
the following assumptions.  Firstly, for simplicity we
consider the case when there is no matter outside the
star so that the conductivity $\sigma=0$ for $r > R$ and
$\sigma \neq 0$ only in a shell with $R_{_{IN}}\le r \le
R$ (e.g. the neutron star crust).  Finally, we require
$\sigma$ to be uniform within this shell (Note that this
might be incorrect in the outermost layers of the neutron
star but is a rather good approximation in the crust as a
whole.).

\section{STATIONARY SOLUTIONS}
\label{srst}
\vskip 0.5 truecm

	In this Section we will look for stationary
solutions of the Maxwell equations, i.e. for solutions in
which the magnetic moment of the magnetic star does not
vary in time. Note that this does not mean that the
electromagnetic fields are independent of time. As a
result of the misalignment between the magnetic dipole
${\mathbf \mu}$ and the angular velocity vector ${\mathbf
\Omega}$, in fact, both the magnetic and the electric
fields will have a {\it periodic} time dependence
produced by the precession of ${\mathbf \mu}$ around
${\mathbf \Omega}$.

	Our strategy in searching for the solution is
based on extending the Deutsch
solution$\footnotesize{^{\cite{d55}}}$ (i.e. the solution
to the Maxwell equations for a misaligned rotating sphere
in a Minkowski spacetime) to a curved spacetime. In
particular, we will distinguish between an exterior
vacuum solution to the Maxwell equations (for which fully
analytic solutions can be found) and an interior
non-vacuum solution. These two solutions will then be
matched at the surface of the star.

	A considerable simplification in the problem
comes from the fact that, at first order in $\Omega$, the
solutions for the electromagnetic fields do not acquire
general relativistic corrections in their angular parts,
and the corresponding Deutsch solution can be used
directly. We will therefore search for a magnetic vector
field with components
\begin{eqnarray}
\label{ansatz_1}
&& B^{\hat r}(r,\theta,\phi,\chi,t) = F(r)
	\left[\cos\chi \cos\theta +
	\sin\chi \sin\theta \cos\lambda(t)\right] \ ,
\\\nonumber\\
\label{ansatz_2}
&& B^{\hat \theta}(r,\theta,\phi,\chi,t) = G(r)
	\left[\cos\chi \sin\theta
	- \sin\chi \cos\theta \cos\lambda(t)\right] \ ,
\\\nonumber\\
\label{ansatz_3}
&& B^{\hat \phi}(r,\theta,\phi,\chi,t) = H(r)
	\sin\chi \sin\lambda(t) \ .
\end{eqnarray}
Imposing now that $J^{\hat r} = J^{\hat\theta} = J^{\hat
\phi} = 0$ (the interior of the star is a perfect
conductor and the exterior of the star is a vacuum), the
radial part of the Maxwell equations
(\ref{max1a}),(\ref{max2b})--(\ref{max2d}) will reduce to
three equations for the radial eigenfunctions $F(r),
G(r)$, and $H(r)$
\begin{eqnarray}
\label{ir_1}
\left(r^2 F\right)_{, r} + 2 e^{\Lambda}r G = 0 \ , &&
\\\nonumber\\
\label{ir_2}
\left(r e^{\Phi} H\right)_{, r}	+ e^{\Phi+\Lambda} F = 0\ , &&
\\ \nonumber\\
\label{ir_3}
H - G = 0  \ . &&
\end{eqnarray}

	Equations (\ref{ir_1})--(\ref{ir_3}) show that,
in the case of infinite conductivity and as far as the
magnetic field is concerned, the use of a slow rotation
metric provides no additional information with respect to
a non-rotating static
metric$\footnotesize{^{\cite{ram00}}}$.

\subsection{Interior Solution}
\label{srst_is}
\vskip 0.5 truecm

	It is important to notice that the system of
equations (\ref{ir_1})--(\ref{ir_3}) combines information
about the structure and physics of the star (through the
metric functions $\Phi$ and $\Lambda$) with information
about the microphysics of the magnetic field (through the
radial eigenfunctions $F$ and $G$). As a result, and as
mentioned in the Introduction, the general relativistic
solution for the interior electromagnetic fields cannot
be given independently of a self-consistent solution of
the Einstein equations for the structure of the star. In
practice, to calculate a generic solution to
(\ref{ir_1})--(\ref{ir_3}), it is necessary to start with
a (realistic) equation of state and obtain a full
solution of the relativistic star. Once the latter is
known, the system of equations (\ref{ir_1})--(\ref{ir_3})
can be solved for a magnetic field which is consistent
with the star's structure and corresponds to a magnetic
configuration of some astrophysical interest. In the case
of a star with constant density, for example, the exact
analytical solution of the inner magnetic field can be
found if a ``stiff matter'' equation of state is used for
the stellar matter$\footnotesize{^{\cite{ram00}}}$.

	Given a dipolar exterior magnetic field, a simple
model for the interior magnetic field comes from assuming
that the latter is continuous through the surface of the
star and down to some inner radius $r=R_{_{IN}}$. The use
of an inner radius removes the problem of suitable
boundary conditions for $r\rightarrow 0$, and reflects
the basic ignorance of the properties of magnetic fields
in the interior regions of neutron stars. In this case,
the radial eigenfunctions $F_{_{IN}}$ and $G_{_{IN}}$ are
obtained through the numerical solution of a system of
coupled ordinary differential
equations$\footnotesize{^{\cite{getal98}}}$. An even
simpler but also cruder model, adopted by Konno and
Kojima$\footnotesize{^{\cite{kk00}}}$ for an aligned
dipole, assumes that the radial eigenfunctions of the
interior magnetic field have the same functional form as
the corresponding eigenfunctions in the exterior of the
star
\begin{equation}
F_{_{IN}}=C_1F_{_{EXT}}\equiv C_1F 
	\hskip 2.0 truecm {\rm and} \hskip 2.0 truecm
G_{_{IN}}=C_1G_{_{EXT}}\equiv C_1G \ ,
\end{equation}
where $C_1$ is a constant to be determined through the
imposition of boundary conditions. This is the model
which we will also consider hereafter as a simple
application of the generic equations derived by Rezzolla,
Ahmedov and Miller$\footnotesize{^{\cite{ram00}}}$

	The form of the internal electric field is
straightforward to derive in the absence of conduction
currents. In this case, in fact, Ohm's laws
(\ref{current1}) and (\ref{current2}) yield the simple
expressions
\begin{eqnarray}
\label{ief_srst_1}
&&E^{\hat r} = \frac{\bar{\omega}r \sin\theta}
	{ce^{\Phi}} B^{\hat \theta}
	= \frac{e^{-(\Phi+\Lambda)}r \sin\theta}{c}
	\bar{\omega}\left(\cos\chi \sin\theta
	- \sin\chi \cos\theta \cos\lambda \right)
	C_1G  \ ,
\\ \nonumber \\
\label{ief_srst_2}
	&&E^{\hat \theta} = -
	\frac{\bar{\omega} r \sin\theta }{ce^{\Phi}} B^{\hat r}
	= - \frac{e^{-\Phi} r \sin\theta }{c} \bar{\omega}
	\left(\cos\chi \cos\theta
	+ \sin\chi \sin\theta \cos\lambda \right)
	C_1 F \ ,
\\ \nonumber \\
\label{ief_srst_3}
	&&E^{\hat \phi} = 0 \ ,
\end{eqnarray}

\noindent where we have taken into account that $\rho_e =
{\mathcal O}(\omega)$ and that the contribution
proportional to $\bar{\omega} \rho_e$ is therefore of
higher order. Note that, apart from the red-shift
correction proportional to $e^{-\Phi}$, equations
(\ref{ief_srst_1})--(\ref{ief_srst_3}) are the same as
those in flat spacetime with $\Omega$ being replaced by
the effective fluid velocity measured by a free falling
observer ${\bar \omega}$.

\subsection{Exterior Solution}
\label{srst_es}
\vskip 0.5 truecm

	The exterior solution for the magnetic field is
simplified by the knowledge of explicit analytic
expressions for the metric functions $\Phi$ and
$\Lambda$. In particular, after defining $N \equiv
e^{\Phi} =e^{-\Lambda}= (1 - 2M/r)^{1/2}$, the system
(\ref{ir_1})--(\ref{ir_3}) can be written as a single,
second-order ordinary differential equation for the
unknown function $F$
\begin{equation}
\label{leg_eq_sim}
\frac{d}{dr}\left[\left(1-\frac{2M}{r}\right)
	\frac{d}{dr}\left(r^2 F\right)\right] - 2F = 0 \ .
\end{equation}
\noindent Let now $x\equiv 1-r/M$, then equation
(\ref{leg_eq_sim}) becomes
\begin{equation}
\label{leg_sol_1}
\frac{d}{dx}\left\{\left(\frac{1+x}{1-x}\right)\frac{d}{dx}
	\left[\left(1-x\right)^2 F\right]\right\}+ 2F = 0 \ ,
\end{equation}
which has an exact solution expressed through the
Legendre functions of the second kind
$Q_{\ell}\footnotesize{^{\cite{j95}}}$. In particular,
the radial eigenfunctions $F(r),\ G(r)$, and $H(r)$, are
found to be 
\begin{eqnarray}
\label{f_of_r}
&& F(r) = - \frac{3}{4M^3}
	\left[\ln N^2 + \frac{2M}{r}\left(1 +  \frac{M}{r}
	\right) \right] \mu
	\ ,
\\\nonumber\\
\label{g_of_r}
&& G(r) = \frac{3 N}{4 M^2 r}
	\left[\frac{r}{M}\ln N^2 +\frac{1}{N^2}+ 1 \right] \mu
	\ ,
\\\nonumber\\
\label{h_of_r}
&& H(r) = G(r) \ .
\end{eqnarray}
Using expressions (\ref{f_of_r})--(\ref{h_of_r}) we can
now determine the value of the matching constant $C_1$ by
requiring that the radial magnetic field is continuous
across the surface of the star, i.e. that $[B^{\hat
r}(r=R)]_{_{IN}} = [B^{\hat r}(r=R)]_{_{EXT}}$. As a
result, we obtain
\begin{equation}
C_1 = 1\ .
\end{equation}

	Collecting all of the expressions for the radial
eigenfunctions, we obtain the following expressions for
the components of the stationary magnetic field
\begin{eqnarray}
\label{sol_mfe_1}
&& B^{\hat r} = - \frac{3}{4M^3}
	\left[\ln N^2 + \frac{2M}{r}\left(1 +  \frac{M}{r}
	\right) \right] (\cos\chi \cos\theta +
	\sin\chi \sin\theta \cos\lambda)\mu
	\ ,
\\\nonumber\\
\label{sol_mfe_2}
&& B^{\hat \theta} = \frac{3 N}{4 M^2 r}
	\left[\frac{r}{M}\ln N^2 +\frac{1}{N^2}+ 1
	\right] (\cos\chi \sin\theta
	- \sin\chi \cos\theta \cos\lambda)\mu
	\ ,
\\\nonumber\\
\label{sol_mfe_3}
&& B^{\hat \phi} = \frac{3 N}{4 M^2 r}
	\left[\frac{r}{M}\ln N^2 +\frac{1}{N^2}+ 1
	\right](\sin\chi \sin\lambda)\mu
	\ .
\end{eqnarray}

	The search for the form of the electric field is
much more involved than for the magnetic field.  Using as
a reference the Deutsch solution, we look for the
simplest solutions of the vacuum Maxwell equations in the
form
\begin{eqnarray}
\label{eef_srst_1}
&& E^{\hat r} = \left(f_1+f_3\right)\left[\cos\chi (3 \cos^2\theta-1)+
	3\sin\chi\cos\lambda
	\sin\theta\cos\theta\right] \ ,
\\
\label{eef_srst_2}
&& E^{\hat \theta} = \frac{\left(f_2+f_4\right)}{2}\Bigg[
	\cos\chi\sin2\theta
	- \left(1-2\sin^2\theta\right)
	\sin\chi\cos\lambda\Bigg] 
	+\left(g_1+g_2\right)\sin\chi\cos\lambda \ ,
\\
\label{eef_srst_3}
&& E^{\hat \phi} = \left[\frac{1}{2}\left(f_2+f_4\right) - 
	\left(g_1+g_2\right)
	\right]\sin\chi\cos\theta\sin\lambda \ ,
\end{eqnarray}
where the unknown eigenfunctions $f_1-f_4$, and $g_1-g_2$
can be found as solutions to the vacuum Maxwell equations
and have radial dependence only. Skipping here the
details of the derivation of the expressions for the
radial eigenfunctions$\footnotesize{^{\cite{kk00}}}$
(whose explicit form is presented in the Appendix) we
here report only the final form of the components of the
stationary vacuum electric field external to the
misaligned magnetized relativistic star
\begin{eqnarray}
\label{eeff_1}
&& E^{\hat r} = 
	\frac{1}{4M^6r^3c}\Bigg\{-\left(
	\frac{5\omega r^3}{8M^4}C_3+\frac{M\Omega}{9R^2}C_2\right)
	\left[6M^4r^3\left(2r-3M\right)
	\ln N^2 + 4M^5r\left(6r^2-3Mr+M^2\right)\right]
\nonumber\\
	&& \ \ \ \ \ \ \ \
	+\frac{3\omega M^3r^4}{2}\ln N^2
	+3M^4\omega r^3\Bigg\}
  \left[\cos\chi (3 \cos^2\theta-1)+3\sin\chi\cos\lambda
	\sin\theta\cos\theta\right]
	\mu \ ,
\\\nonumber \\
\label{eeff_2}
&& E^{\hat \theta} = - 
	\frac{3}{2M^6r^4Nc}\Bigg\{-\left(
	\frac{5\omega r^3}{8M^4}C_3+\frac{M\Omega}{9R^2}C_2\right)
\nonumber\\
	&& \ \ \ \ \ \ \ \ \ \
	\left[3M^4r^3\left(r^2-3Mr+2M^2\right)
	\ln N^2 + 2M^5r^2\left(3r^2-6Mr+M^2\right)\right]
	+\frac{\omega r^3M^2}{2}\Bigg\}
\nonumber\\
	&& \ \ \ \ \ \ \	
	\times \left[2\cos\chi\sin\theta\cos\theta -
	\left(\cos^2\theta-\sin^2\theta\right)
	\sin\chi\cos\lambda\right] \mu
\nonumber\\ \nonumber\\
	&& \ \ \ \ \ \ \ \
	+\frac{3\bar{\omega} r}{8 M^3 c N}\left[\ln N^2+
	\frac{2M}{r}\left(1+\frac{M}{r}\right)\right]
	(\sin\chi\cos\lambda)
	\mu \ ,
\\\nonumber \\
\label{eeff_3}
&& E^{\hat \phi} = - \Bigg\{
	\frac{3}{2M^6r^4Nc}\Bigg\{-\left(
	\frac{5\omega r^3}{8M^4}C_3+\frac{M\Omega}{9R^2}C_2\right)
\nonumber\\
	&& \ \ \ \ \ \ \ \ \ \
	\left[3M^4r^3\left(r^2-3Mr+2M^2\right)
	\ln N^2 + 2M^5r^2\left(3r^2-6Mr+M^2\right)\right]
	+\frac{\omega r^3M^2}{2}\Bigg\}
\nonumber\\
	&& \ \ \ \ \ \ \ \ \ \ \ \
	- \frac{3\bar{\omega} r}{8 M^3 c N}\left[\ln N^2+
	\frac{2M}{r}\left(1+\frac{M}{r}\right)\right]
	\Bigg\}	(\sin\chi\cos\theta\sin\lambda)
	\mu \ ,
\end{eqnarray}
where $C_2$ and $C_3$ are arbitrary constants to be
determined through the imposition of boundary conditions.
	
	Expressions (\ref{eeff_1})--(\ref{eeff_3})
confirm that the general relativistic dragging of
reference frames introduces a new contribution to the
form of the electric field which does not have a flat
spacetime analogue. This effect is ${\mathcal O}(\omega)$
and therefore is present already in our first-order slow
rotation approximation. This is in contrast with what
happens for the magnetic fields, where higher order
approximations of the form of the metric are necessary
for frame dragging corrections to appear.

	The values of $C_2$ and $C_3$ can be found after
imposing the continuity of the tangential electric field
across the stellar surface. Using
(\ref{ief_srst_1})--(\ref{ief_srst_3}) as solutions for
the internal electric field and imposing that $[E^{\hat
\theta}(r=R)]_{_{IN}}=[E^{\hat \theta}(r=R)]_{_{EXT}}$,
as well as $[E^{\hat \phi}(r=R)]_{_{IN}}=[E^{\hat
\phi}(r=R)]_{_{EXT}}$, yields
\begin{eqnarray}
\label{c2}
C_2=\frac{9R^5\left[\ln N^2_{_R} + 2M/R 
	\left(1+M/R\right)\right]}{8M^3\left(3R^2-6MR+M^2\right)
	+12M^2R\left(R^2-3MR+2M^2\right)\ln N^2_{_R}}\ ,
\\ \nonumber \\
C_3=-\frac{2M^3\left[\ln N^2_{_R} + 2M/R\right]}
	{30MR^2-60M^2R+10M^3+\left(15R^3-45MR^2+
	30M^2R\right)\ln N^2_{_R}}\ ,
\end{eqnarray}
with $N^2_{_R} \equiv N^2(r=R) = 1 - 2M/R$.

\noindent Note that in the limit of an aligned dipole
($\chi=0$), equations (\ref{eeff_1})--(\ref{eeff_2})
reduce, after some algebraic manipulations, to the vacuum
solutions found by Konno and
Kojima$\footnotesize{^{\cite{kk00}}}$.
\begin{eqnarray}
\label{efkk_1}
&& E^{\hat r} = 
	\frac{1}{4M^6r^3c}\Bigg\{C
	\left[6M^4r^3\left(2r-3M\right)
	\ln N^2 + 4M^5r\left(6r^2-3Mr+M^2\right)\right]
\nonumber\\
	&& \ \ \ \ \ \ \ \ \ \
	+\frac{3\omega M^3r^4}{2}\ln N^2
	+3M^4\omega r^3\Bigg\}(3 \cos^2\theta-1)
	\mu \ ,
\\\nonumber \\
\label{efkk_2}
&& E^{\hat \theta} = - 
	\frac{3}{M^6r^4Nc}\Bigg\{C 
	\left[3M^4r^3\left(r^2-3Mr+2M^2\right)
	\ln N^2 + 2M^5r^2\left(3r^2-6Mr+M^2\right)\right]
\nonumber\\
	&& \ \ \ \ \ \ \		
	+\frac{\omega r^3M^2}{2}\Bigg\}
	\sin\theta\cos\theta \mu \ ,
\end{eqnarray}
where the constant $C$ is related to the integration
constant $c_2$ given by formula (2.11) of Konno and
Kojima's paper$\footnotesize{^{\cite{kk00}}}$ by
\begin{equation}
C=-\left(\frac{5\omega r^3}{8M^4}C_3+\frac{M\Omega}{9R^2}C_2\right)
=\frac{c_2}{\mu}-\frac{\omega r^3}{M^4} .
\end{equation}

\section{CONCLUSION}
\label{conclusion}
\vskip 0.5 truecm

	In a recent related
work$\footnotesize{^{\cite{ram00}}}$, we have presented
analytic general relativistic expressions for the
stationary electromagnetic fields in the crust and vacuum
region of a slowly-rotating magnetized neutron
star. There, the star was considered isolated and in
vacuum, but no assumption was made about the orientation
of the dipolar magnetic field with respect to the
rotation axis. As an application of the general formalism
developed, we then found an exact solution for the
interior magnetic field for a star with uniform density
and a stiff matter equation of state.

	In this paper we have instead considered the case
of an internal magnetic field with a radial dependence
which is the same as that of the external dipolar
magnetic field. Strictly speaking this cannot be entirely
correct since the interior magnetic field cannot be
determined independently of the structure of the rotating
star. However, this remains an interesting application of
our equations to a case which has a Newtonian analogue
and has recently been discussed in the
literature$\footnotesize{^{\cite{kk00}}}$.

	The solutions presented in this paper provide a
lowest order analytic form for the electromagnetic fields
in the spacetime of a slowly rotating misaligned dipole
and agree, in the case of an aligned dipole, with the
results of Konno and
Kojima$\footnotesize{^{\cite{kk00}}}$. While our
solutions have been found with some simplifying
assumptions, they also allow the major features of a
realistic solution to be seen and could therefore be used
in a variety of astrophysical situations.

\section*{ACKNOWLEDGMENTS}
\vskip 0.5 truecm

Financial support for this work is provided by the
Italian Ministero dell'Universit\`a e della Ricerca
Scientifica e Tecnologica. J.C.M. also acknowledges
partial support from the INFN. B.A. is grateful to the
INFN for supporting the visit to SISSA where part of this
research was carried out.

\section*{APPENDIX}
\vskip 0.5 truecm

	For completeness, we here report the explicit
forms of the functions $f_1-f_4$, and $g_1,\;g_2$ used in
expressions~(\ref{eef_srst_1})--(\ref{eef_srst_3}) for
the components of the external electric field

\begin{eqnarray}
\label{fs_sol}
&& f_1 = \frac{\Omega}{6 c R^2} C_2
	\left[\left(3-\frac{2r}{M}\right)
	\ln N^2 + \frac{2M^2}{3r^2}+\frac{2M}{r}-4\right]
	\mu \ ,
\\\nonumber\\
&& f_2 = -\frac{\Omega}{c R^2} C_2 N
	\left[\left(1-\frac{r}{M}
	\right)\ln N^2-2-\frac{2M^2}{3r^2N^2}\right]
	\mu \ ,
\\\nonumber\\
&& f_3 = \frac{15\omega r^3}{16M^5c}\Bigg\{C_3\left[
	\left(3-\frac{2r}{M}\right)
	\ln N^2 + \frac{2M^2}{3r^2}+\frac{2M}{r}-4\right]
%	\nonumber\\
%	&& \hskip 4.0cm
	+\frac{2M^2}{5r^2}\ln N^2
	+\frac{4M^3}{5r^3} \Bigg\}\mu \ ,
\\\nonumber\\
&& f_4 = -\frac{45\omega r^3}{8M^5 c}N\left\{C_3
	\left[\left(1-\frac{r}{M}
	\right)\ln N^2-2-\frac{2M^2}{3r^2N^2}\right]
	+\frac{4M^4}{15r^4 N^2}\right\}\mu \ ,
\\\nonumber\\
&& g_1 = \frac{3\Omega r}{8 M^3cN}\left[\ln N^2+
	\frac{2M}{r}\left(1+\frac{M}{r}\right)\right]
	\mu \ ,
\\\nonumber \\
\label{g4_sol}
&& g_2 = -\frac{\omega}{\Omega}g_2
	= -\frac{3\omega r}{8 M^3cN}\left[\ln N^2+
	\frac{2M}{r}\left(1+\frac{M}{r}\right)\right]
	\mu \ ,
\end{eqnarray}
where $g_1+g_2 = ({\bar \omega}/\Omega) g_1$.

\end{document}